\documentstyle[11pt,newpasp,twoside,epsf,psfig]{article}

\begin{document}

\title{Sinks of Light Elements in Stars - Part II}

\author{M.H. Pinsonneault}
\affil {The Ohio State University, Department of Astronomy, 
140 W. 18th Ave. Columbus, Ohio 43210, USA}

\author {C. Charbonnel}
\affil{Laboratoire d'Astrophysique de l'Observatoire Midi-Pyr\'en\'ees, CNRS-UMR
5572, 14, av.E.Belin, F-31400 Toulouse, France}

\author{C.P. Deliyannis}
%CC1906 Complete address
\affil{Indiana University, Astronomy Department,
319 Swain Hall West, 727 E. 3rd Street, Bloomington, IN 47405-7105, USA}

\begin{abstract}

See the abstract given in Part I (Deliyannis, Pinsonneault,
Charbonnel, hereafter DPC, in these proceedings).  In Part II we
discuss the lithium data for metal-poor stars and the constraints it
places on stellar depletion.  There are a variety of indicators that
place interesting bounds on the degree of stellar depletion, and in
contrast to the other two sections the data for Population II stars
provides weaker evidence for stellar lithium depletion.  We review
both the theoretical studies and the observational data, and
critically evaluate the degree of stellar depletion consistent with
the data.

\end{abstract}

\section{Introduction}

The study of lithium in metal-poor stars has implications for stellar
structure, galactic chemical evolution, and Big Bang nucleosynthesis
(BBN).  There has thus been considerable observational and theoretical
effort on this question, and we will therefore attempt to draw
together the main themes and outstanding questions rather than
undertake a detailed analysis of the fine points.  This paper will be
organized as follows. 
In section 2, we recall the main trends of the theoretical expectations. 
In section 3, we discuss the major features of the observational data; 
in particular we stress the similarities and differences with the population I 
pattern.  
Section 4 is devoted to the comparison of different classes of theoretical 
models with 
the data, and our conclusions are summarized in section 5.

\section{Main Trends of the Theoretical Expectations}

Let us first briefly recall that the so-called standard case refers 
to models which exclude any kind of transport processes of the chemicals 
in the radiative zones, and thus consider only convection as a 
mixing mechanism.  Lithium is destroyed at
moderate temperature by stellar interior standards; 
for typical 
%CC1906 I add main sequence here so that it is defined.
main sequence (MS) and pre-MS densities, Li${^6}$ burns at a
time scale comparable to or shorter than the evolutionary time scale at
around 2 million K and Li${^7}$ burns at around 2.6 million K.  Both
beryllium and boron are less fragile, with characteristic burning
temperatures of order 3.5 million K and 5 million K respectively.
Because the observational data points to modest lithium depletion in
halo stars (and even this conclusion is controversial!) we will
restrict our discussion of light element depletion to lithium.

The major predictions of classical 
%CC1906 Better to define classical now, and remove this definition 
% from the next paragraph
(sometimes referred to as standard) stellar evolution models for halo
stars are summarized in Deliyannis et al. (1990). 
They depend mainly on the variations of the depth of the stellar convective 
envelope (and thus of the temperature at its base) 
with the stellar mass, metallicity and evolutionary stage. 
Pre-MS depletion increases with decreased mass, and there will
therefore be a strong
decrease of lithium with decreased $T_{eff}$ for cool stars.  Main
sequence depletion is predicted to be minimal for all but the coolest
stars; the absolute degree of lithium depletion decreases with
decreased metal abundance.  The net effect predicted is that hot halo
dwarfs should exhibit little or no dependence of their surface lithium
abundance on effective temperature, and for the lowest metal
abundances there should be a minimal dependence on [Fe/H], in the
sense that lower abundances would be predicted for higher
metallicities.  This implies a small dispersion in lithium at fixed
effective temperature; in the Population I case, it also implies a
weak dependence of lithium on age which can be tested in open
clusters.

Classical 
%CC1906 now defined above : (sometimes referred to as standard)  
stellar models neglect
some physical processes which are known to be important for
interpreting the surface lithium abundances of stars.
The linked phenomena of gravitational settling and thermal diffusion,
which are solidly based in our knowledge of plasma physics,
are among the most important. 
Atomic diffusion is a fundamental process which 
must occur in the stellar gas unless some macroscopic 
motions counteract it, 
causing heavy elements to sink with respect to light ones under the
conditions applicable for halo dwarfs.  The time scale for this
process decreases as the depth of the surface convection zone decreases.

Theoretical models which include 
%CC1906 I suggest to add a footnote here to define what we mean by 
%"pure atomic diffusion"
pure atomic diffusion\footnote{By "pure atomic diffusion" we mean that
it is not counteracted by any macroscopic process in the stellar
radiative zones}
%CC1906 
(see Michaud et al. 1984 for the first computations for halo stars) 
therefore predict that lithium sinks below the surface
convection zone for the conditions appropriate for subdwarfs,
and furthermore that the degree of diffusion increases with increased 
$T_{eff}$.  Microscopic diffusion will not generate a dispersion in
abundance at fixed $T_{eff}$, and the effects at a given surface
temperature are not strongly metallicity dependent.

Stellar winds can counteract and prevent atomic diffusion without leading to 
nuclear destruction (Vauclair \& Charbonnel 1995). The corresponding mass loss 
necessary in the hottest halo stars is in excess of the value inferred from an 
extrapolation of the solar Mdot to halo stars by a factor of about 10 to 30, 
but not by a degree that can be ruled out observationally. 
For even larger mass loss rates the outer layers containing lithium
can be removed; as noted by Swenson \& Faulkner (1992) the finite
depth of the surface convection zone must be accounted for and models with
%CC1906 
strong (stronger than inferred by Vauclair \& Charbonnel 1995)
%CC1906 
mass loss alone are incompatible with the Population I lithium pattern.
For hot halo stars the combined effect of diffusion and mass loss 
produce both lithium depletion and a 
%CC1906
small 
%CC1906
dispersion in abundance.

There are known mechanisms for mild mixing in the
radiative envelopes of low mass stars.  The two most frequently
studied are rotationally-induced mixing and turbulence induced by
gravity waves (see Pinsonneault 1997 for a review
%CC1906
and DPC).
%CC1906
Gravity waves can
be produced by turbulence in the surface convection zone; because the
convection zone depth is a strong function of mass, this can produce
mass-dependent lithium depletion that is a function of time on the
main sequence.  It would not produce a dispersion in abundance for a
sample of uniform age and composition, but abundance differences could
be generated by a range of age and composition as seen in field halo
stars.

The degree of rotational mixing depends on several major factors.  
Low mass Population I stars are observed to have a range of surface rotation 
rates
and stars of the same mass, composition, and age could therefore have
different initial angular momenta, different rotation rates as a
function of time, and different degrees of rotational mixing.  
A dispersion in surface lithium abundance, even at fixed effective
temperature in clusters, is therefore expected.  However, the observed
distribution of rotation rates in young clusters is nongaussian which
strongly affects the expected lithium depletion pattern.  We
also cannot directly observe the initial conditions for Population II
stars. To predict the detailed distribution of abundances 
the best we can to is to infer the distribution of initial conditions from young
open cluster stars (see Pinsonneault et al. 1999, hereafter PWSN).
The degree of mixing is also directly linked to the internal transport
of angular momentum (e.g., Zahn 1992, Maeder 1995). 
Rotational mixing can also be inhibited by gradients in
mean molecular weight - induced by nuclear burning in the cores of
stars and possibly also by gravitational settling of helium in their
outer layers (see also Michaud and Vauclair in these proceedings). 
In contrast with classical models, more modern models including mild 
mixing below the envelope on the main sequence
can simultaneously produce modest depletions of species, such as
lithium and beryllium, that burn at very different temperatures.
Rotational mixing will also be a function of age, and some classes of
models predict trends with [Fe/H] and effective temperature that can
be tested in halo stars.

Realistic models should include the possible interactions between the 
above, since mass loss can counteract atomic diffusion and diffusion 
can interact with mixing (also see Michaud, these proceedings).  We note
that Vauclair (these proceedings) has proposed a nonlinear interaction
between mixing and microscopic diffusion which would permit negligible 
halo star lithium depletion; note, however, that the details of such
an interaction need to be computed and that such a cancellation would
still have to be consistent with the globular cluster and Population I
star data.   

\section{Observational Pattern}

As we have just seen the surface lithium abundances of stars are sensitive 
to a variety of effects, both those accounted for in classical stellar models 
and those caused by physically well-motivated but still so-called
non-standard effects. 
Uniqueness is thus a real issue when interpreting the observational data. 
We will therefore begin with the overall conclusions from studies of 
Population I stars, and then proceed to the current status of the 
observational data for Population II stars.

\subsection{Population I Properties}

The properties of Population I stars are summarized in DPC. Here we briefly 
recall those that are important for the problem of lithium depletion in 
Population II stars.  In progressively
older open cluster stars there is clear evidence for increased lithium
depletion with age and a dispersion in abundance at fixed $T_{eff}$
which is inconsistent with classical models.  The predicted dropoff of
lithium for cool stars from pre-MS burning is clearly seen.  The
overall properties favor mild mixing below the envelope on the MS; 
there is also evidence for microscopic diffusion playing a role for F stars 
and in the helioseismic inversions of the solar sound speed relative to
theoretical models (see Guzik \& Cox 1993, 
Richard et al. 1996, Basu et al. 2000).  
Large enough amounts of mass loss to directly cause lithium depletion are
inconsistent with the observed population I pattern (Swenson \&
Faulkner 1992).  As of this time a single theoretical model capable of
explaining all of the Population I data has not yet been found (Talon
\& Charbonnel 1998, PWSN).

\subsection{Overall Population II Properties}

Beginning with the pioneering work of Spite \& Spite (1982),
there have been a series of progressively more sophisticated
observational studies of lithium in halo field stars; the largest
sample is that of Thorburn (1994).  Ryan et al. (1999, hereafter RNB) 
obtained a smaller sample with a lower formal error ($\sigma$
$\sim$ 0.036 dex) than the errors in earlier studies 
($\sigma$ $\sim$ 0.07-0.09 dex). 
There have also been preliminary studies of small samples of stars
near the turnoff in globular clusters.  Different investigators of
field stars agree on some general properties:

1.  Halo stars hotter than 5800 K exhibit a weaker dependence on $T_{eff}$,
    [Fe/H], and a smaller dispersion than seen in Population I stars.
    There is vigorous debate about the existence and magnitude of any
dispersion in the field star case, and there are active controversies
about trends with $T_{eff}$ and [Fe/H].

2.  Turnoff globular cluster stars were first studied by Molaro \&
    Pasquini (1994), who reported a Li abundance for a turnoff star in
    NGC 6397 consistent with the halo plateau abundances.  These are 
    technically challenging observations owing to the faintness of the 
stars and the need for
    high resolution spectroscopy.  A subsequent Keck study of the
    globular cluster M92 (Deliyannis et al. 1995, Boesgaard et al. 1998)
    revealed a large scatter, similar in morphology to the old open
    cluster M67.  The sample, however, is small (seven stars,
    including only three with S/N greater than 40).  Pasquini \&
    Molaro (1997) also observed a range of Li abundances in three
    turnoff stars in the intermediate metal abundance globular cluster
    47 Tuc.  Any successful
    theory must explain both the cluster and field star patterns; more
    data is clearly needed for the globular cluster stars 
(see also Part III, Charbonnel, Deliyannis \& Pinsonneault).

We now turn to a summary of the most recent data on the important
global features in field halo stars.

\subsection{Trends with $T_{eff}$}

There has been a spirited debate about the existence and magnitude of
trends in the halo star data with metallicity and effective
temperature.  The slope with metal abundance is important for
constraining the galactic chemical evolution contribution to the
observed lithium abundances, and it may also contribute to the small
scatter in the data at fixed effective temperature.  Trends with
effective temperature are important as a diagnostic of the mass
dependence of any physical processes which affect the surface
abundances.  Thorburn (1994) reported evidence for a positive slope of
lithium with respect to $T_{eff}$; this conclusion was
challenged by Molaro et al. (1995).  Subsequently Ryan et al. 
(1996) reanalyzed their data, claiming confirmation of the original results; 
see also Bonifacio \& Molaro (1997).  The existence of a modest 
rising trend with increased T$_{eff}$ is only predicted in the models 
including atomic diffusion and stellar winds (Vauclair \& Charbonnel 1995). 
However, what is even more important than the existence of a mild mean trend 
is the thing which is {\it not} seen : 
any evidence for a decline in lithium among the hottest stars.

As discussed above, models which include only microscopic diffusion predict
a decline in surface lithium for the hottest halo stars; models with
strong depletion from mixing also predict a downwards trend in lithium
for the hottest stars (Chaboyer \& Demarque 1994). This observational
fact therefore constitutes an important limit on lithium depletion in
halo stars.

\subsection{Trends with [Fe/H]}

The existence of trends of lithium with metallicity is a signature of
galactic chemical evolution, and it could also contribute to the
dispersion observed in halo stars.  The majority of the observational 
investigations have looked for a correlation between [Li] and 
[Fe/H]; in parallel to the controversy over trends with effective
temperature, there have been conflicting results on metallicity
trends.  Both Thorburn (1994) and Ryan et al. (1996) found some
evidence for an increase in [Li] with [Fe/H] with a slope of order
0.1.  This was disputed by Molaro et al. (1995) and Bonifacio \&
Molaro (1997).  RNB obtained a sample with a small intrinsic range in
effective temperature, but a wider range in metallicity.  They could
therefore evaluate metallicity but not effective temperature trends,
and found a slope of [Li] with respect to [Fe/H] consistent with the
Thorburn (1994) level.  Chemical evolution trends should most
logically be evaluated in the linear Li - linear Fe/H plane (see
Olive and Matteucci, these proceedings). 

A general feature of the derived chemical evolution trends is that 
they are sensitive to the source used for the metallicity, 
the subset of the data which is used, and the treatment
of outliers in the fit.  Ryan et al. (1996) also noted that the
evidence for trends in the data with $T_{eff}$ and [Fe/H] is more
convincing in a bivariate analysis than when either variable is
treated separately.  The existence and magnitude of metallicity
trends is also important for the interpretation of the dispersion in
abundance (see below).  The metallicity dependence of any rotational
mixing is small (PWSN) and would be difficult to disentangle from
chemical evolution effects.

\subsection{Dispersion}

The dispersion in the lithium abundances of halo plateau stars has
been the subject of a number of studies.  This is largely because the
existence or absence of a detectable range in abundance at fixed
metallicity and $T_{eff}$ is the best direct test and constraint on 
the transport processes of chemicals in these stars. Lithium abundances can be
studied as a function of age in the Population I case, which makes it
easier to unambiguously distinguish between different classes of
theoretical models (or at least rule bad models out).  All of the
metal-poor stars that we observe are old, and we therefore cannot
directly reconstruct the depletion of lithium by sorting stars of
progressively increased age into an evolutionary sequence.

Studies of the dispersion tend to fall into two groups.  Some
investigators (Deliyannis et al. 1993, Thorburn 1994) found evidence for a
dispersion at a low level; others (Spite et al. 1996, Bonifacio \&
Molaro 1997) placed bounds on the dispersion consistent with their
observational errors.  The level of dispersion inferred by Deliyannis
et al. (1993) is not inconsistent with the latter two studies (greater
than 0.04 dex as compared with less than 0.08 and 0.07 dex
respectively). In a recent paper, RNB have claimed a more stringent
constraint on the overall dispersion.  We examine this most recent
data set below; a comparison of models including rotational mixing
with the Thorburn (1994) data set was performed by PWSN.

The formal dispersion of the RNB data set is 0.053 dex, greater than their
observational error of 0.036\footnote{(not 0.033 as noted in the paper)}.
They attribute this to a correlation between metallicity and lithium
abundance, e.g. chemical evolution rather than stellar depletion.
There is a substantial overlap between the RNB and Thorburn (1994)
data sets, and the markedly lower dispersion inferred by RNB can be
traced directly to differences in equivalent width measurements.  In
Figure 1 we illustrate and compare the properties of the RNB sample
(excluding one upper limit) with the stars in common as measured by
Thorburn (1994); both have been shifted to the same effective
temperature scale.  There are both significant zero-point shifts and a
marked difference in the overall dispersion of the sample.  RNB
attribute this to possible scattered light and sky subtraction issues
in the 
%CC1906 Thorrburn (1994) 
Thorburn (1994) data set.  We
note, however, that similar differences appear in samples in common
with other investigators\footnote{(e.g. compare G64-12 and CD -33 1173 as
measured by both Thorburn 1994 and Spite \& Spite 1993 to their
relative abundances in RNB)}.
The systematic differences between various observational data sets
therefore require more scrutiny, especially given the relatively small 
sample size of the RNB data set and the small number of overdepleted
stars expected for modest stellar depletion.

\begin{figure}[h] 
\epsfxsize=9cm
\epsfysize=9cm
\hskip 1in 
\epsfbox{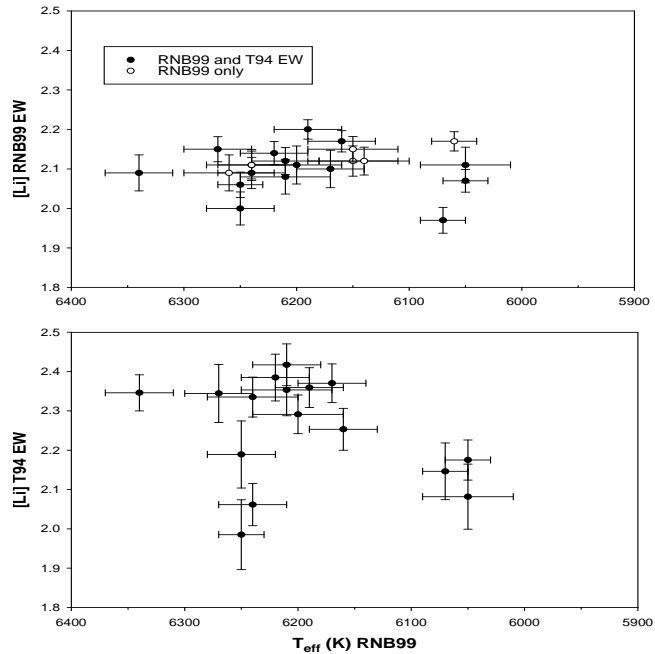}
\caption{Data from RNB (top panel) compared with data for stars in
common with Thorburn (1994) corrected to the same effective temperature
(bottom panel)}
\end{figure}

%\begin{figure}
%\centerline{
%\epsfxsize=\hsize
%\epsfbox[18 144 592 738]{IAU198fig1.eps}
%}
%\end{figure}

We compare the theoretical distribution of the lowest depletion case
of PWSN to the RNB data in Figure 2.  The majority of young low mass 
stars in open clusters are slow rotators, which implies that 
they should have experienced similar rotational histories and similar
degrees of rotational mixing.  However about 1/5
of the stars in open clusters are observed to be rapid rotators and
these should manifest themselves as overdepleted objects, producing an
excess dispersion which is measureable. 
The existence of a core in the sample with minimal internal
dispersion therefore does not by itself rule out more modest stellar
destruction (as noted by RNB).  The existence and number of outliers
is a more stringent test.  In the raw RNB sample there
are three stars more than 0.1 dex below the median, one of which has
an upper limit of 1.36 for its abundance; this simulation would
predict ~4 depending on the criterion for defining what constitutes an
overdepleted star.  It is legitimate to question whether the highly
overdepleted star is produced by the same mechanism as the other
stars, but in any case the sample size is small and it is certainly
difficult to make a persuasive case against modest
depletion factors based on the data without chemical evolution corrections.

\begin{figure}[h] 
\epsfxsize=9cm
\epsfysize=9cm
\hskip 1in 
\epsfbox{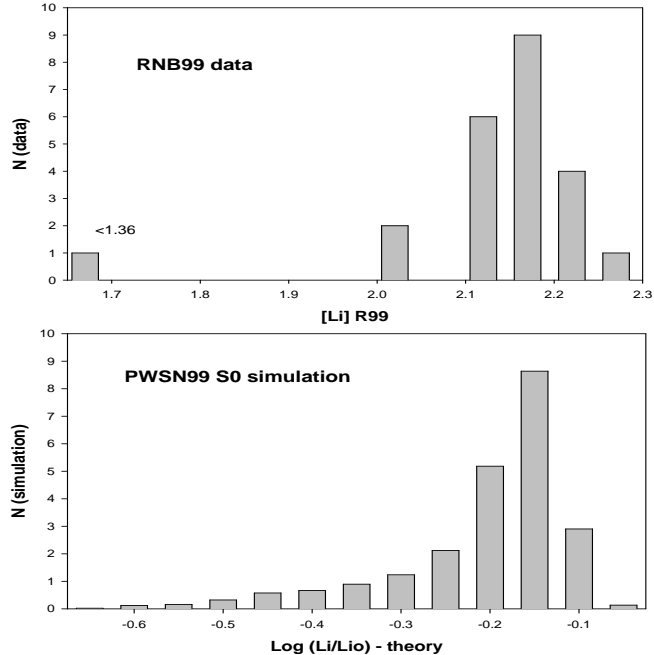}
\caption{Data from RNB (top panel) compared with the theoretical
simulation of PWSN for their low depletion case.  The theoretical
distribution has been convolved with an observational error of 0.03 dex.}
\end{figure}

RNB placed more stringent constraints than the above based upon
attributing some of their small dispersion to galactic chemical
evolution.  RNB fitted the data for a trend with [Fe/H] and concluded
that there was a 10 \% probability that as few outliers (one) as
observed would be present by chance.  This conclusion depends on the
usage of a logarithmic, rather than a linear, relationship between
lithium abundance and metallicity (Pinsonneault et al. 2000).  In
conclusion, the RNB data places more severe constraints on the
dispersion in abundance than previous studies, and depending on the
treatment of trends with metal abundance it may either be consistent
with modest stellar depletion factors or places a bound of order 0.1
dex on the absolute depletion from the class of rotational mixing
models considered by PWSN.

The existence of stars above the plateau may also provide some
important clues; they could either be underdepleted or they could have
experienced lithium production.  Stars with very low initial angular
momentum would experience much less rotational mixing than the norm and would
therefore appear as underdepleted.  However, there are strong observational
selection effects against detecting very slow rotators in open
clusters, and it is therefore difficult to estimate the
fraction of such objects that would be expected in rotational mixing
models.  An alternative explanation would be differential lithium
production.  King et al. (1996) examined the most prominent such star,
BD+23:3912, and found no evidence for lithium production in the
abundances of other elements that would be affected by the main
mechanisms (see King et al. 1996 for a detailed discussion and caveats).

%CC1906 Shouldn't we change all $Li{^7}$ in ${^7}$Li?
\subsection{$Li{^6}/Li{^7}$}

Li$^6$ is more fragile than Li$^7$
and it is not produced in significant quantities in standard BBN
models.  The detection of Li$^6$ in halo stars can therefore be used
to set powerful constraints on the absolute depletion of Li$^7$, with
the caveat that the initial Li$^6$ abundance must be inferred from
chemical evolution models.  Smith et al. (1993) first claimed a
detection of Li$^6$ in the halo star HD 84937. 
This was confirmed in subsequent studies by different investigators who 
added two more possible detections and a number of upper
limits (see Cayrel et al. 1999, Hobbs et al. 1999, Nissen et al. 1999
for recent work on the subject and Nissen in these proceedings). 
The detected amount of Li$^6$ is
small, but it appears to be secure. The amount is greater than would
be expected from the beryllium and boron data, suggesting that
alpha-alpha fusion may contribute to the production of Li$^6$.

One important uncertainty in the usage of Li$^6$ data is therefore what
the initial abundance of the species could be; for example, Lemoine et
al. (1996) and Cayrel et al. (1999) obtained bounds of a factor of four
and three respectively on the absolute depletion of Li$^6$ in HD 84937.  
PWSN argued that an even higher initial abundance could not be
excluded, and considered the extreme limiting case where the halo Li$^6$
abundance could have been as high as the solar system value.

The second uncertainty is the ratio of Li$^6$ to Li$^7$ depletion.
Nuclear burning in the convective envelope 
in standard models would produce strong Li$^6$
depletion before any Li$^7$ depletion occurred; this has been used to
argue that any detected Li$^6$ implies negligible $Li^7$ depletion
(e.g. Lemoine et al. 1996).  Both models with microscopic diffusion
and models with mild envelope mixing, however, predict simultaneous
detectable depletion of both isotopes to varying degrees, with Li$^6$
being more sensitive but not infinitely so (see PWSN).  Nonetheless,
the Li$^6$ data does provide one of the best independent checks on any
stellar depletion of Li$^7$, and it indicates that large depletion
factors are very unlikely to be consistent with the observed
detections (see below.)  

\section{Theoretical Constraints on Lithium Depletion}

In light of the observational data above, what can we infer about the
depletion of lithium in halo stars?  The first and most generally
agreed-upon conclusion is that a variety of observational tests make a
large depletion factor unlikely.  There have been three major features
of the halo data which have been used to constrain the absolute
depletion: the degree of dispersion in the halo plateau, the detection
of $Li^6$ in some halo stars, and the absence of a decline in surface
Li for hotter halo stars. 

\subsection{Bounds from the Dispersion in the Plateau}

PWSN inferred a range of 0.2-0.4 dex
depletion factors from models including rotational mixing; the lower
end of the range was more consistent with the dispersion inferred from
the Thorburn (1994) data set, while the upper end of the range
permitted the rare highly overdepleted stars to be explained within
the framework of rotational mixing.  As discussed above, the most
recent data set of RNB is marginally consistent with the lower end of
the depletion range in PWSN (of order 0.2 dex.)  Other investigators
(e.g. Bonifacio \& Molaro 1997, RNB)
have claimed more stringent limits of order 0.1 dex on the absolute 
depletion based
upon the small (or, in their view, nonexistent!) dispersion in the
halo Li data.  We will return to this claim after reviewing other
measures based upon the detection of $Li^6$ and the absence of the
observed signature of pure microscopic diffusion in halo stars.

\subsection{Bounds from $Li^6$ / $Li^7$ Measurements}

PWSN set a less severe, but firm, limit of
0.5-0.6 dex $Li^7$ depletion from the measured $Li^6$ / $Li^7$
abundance ratios in halo stars under the assumption that the halo
stars did not have a $Li^6$ abundance higher than the solar system
value.  Lemoine et al. (1996) derived a bound of a factor of 4 on the
absolute $Li^6$ depletion of HD 84937, which in the rotationally mixed
models of PWSN would imply a bound of ~0.25 dex on the $Li^7$
depletion, while Cayrel et al. (1999) used the lithium data in the
same star to set a bound of 0.1 dex on its $Li^7$ depletion; both of
these calculations, however, are dependent on the chemical evolution
model which is used.  We note that if the Cayrel et al. (1999) bound
of a factor of three $Li^6$ depletion is used in conjunction with the
PWSN models, an absolute $Li^7$ depletion of ~0.15 dex is inferred for
HD 84937 rather than an upper limit of 0.1 dex.

\subsection{Bounds from the Flatness of the Plateau}

Vauclair \& Charbonnel (1998) used a different set of properties to
infer a stellar depletion factor.  
%CC1906 I suggest to omit "The inclusion" here. 
Pure microscopic diffusion in halo models would produce a strong decrease 
in surface lithium with increased $T_{eff}$ which is not observed. 
It is therefore clear that something must be inhibiting microscopic
diffusion, especially given the improved agreement with
helioseismology from the inclusion of micrscopic diffusion in solar
model calculations.  They noted that there is a subsurface peak in the
$Li^7$ abundance of the pure diffusion models 
which does not vary greatly across the plateau.

Vauclair \& Charbonnel (1995) (see also Swenson 1995) argued that mass loss 
at a rate 10-30 times greater than the solar value could counteract the effects
of diffusion if the rate was tuned across the lithium plateau. This
would have the effect of exposing a uniform abundance across the
plateau, with an absolute depletion of 0.15 dex.  Vauclair \&
Charbonnel (1998) noted that in the presence of sufficiently mild
mixing the height of this peak could be preserved, implying that a
uniform depletion of order 0.15 dex could apply if the absence of a
measurable surface signature of diffusion arose from either the
competing effects of mass loss or the interaction of diffusion and
mild mixing.  As noted by Chaboyer \& Demarque (1994), sufficiently
strong mixing can cancel the effects of diffusion while not preserving
the height of the peak; however, models with the high degree of
depletion inferred by that 
%CC1906
latter 
%CC1906
paper are difficult to reconcile with the
other observational tests of lithium depletion.

\subsection{Is There Any Depletion?}

In light of the remarkable observed properties of the halo lithium
plateau, it is reasonable to ask whether there is in fact any
depletion at all.  In other words, are we directly seeing the
primordial lithium abundance (e.g. Bonifacio \& Molaro 1997) or is
the primordial lithium abundance in fact {\it lower} than the observed
values because of a significant contribution from galactic chemical
evolution (RNB)?

``Standard'' stellar models are sometimes invoked as evidence against
significant depletion.  These classical models, however, achieve this
prediction by simply neglecting known physics rather than by
demonstrating that it is unimportant.  For example, one could
construct stellar models which ignore the CNO cycle, and they might
even agree with some data, but it does not follow that these should be
placed on an equal physical basis with models that include the known
nuclear physics.
In particular, atomic diffusion
cannot be excluded arbitrarily from the computations, on the pretext 
that it produces unobserved features. This disagreement is a simple 
signature of some macroscopic motions (mass loss or rotation-induced 
motions) which counteract the diffusion process. 

Any model predicting zero stellar depletion in halo field stars
must be reconciled with the apparent
% model must be reconciled with the apparent
scatter in the globular cluster
turnoff stars.  If both the halo stars and the globular cluster stars
are depleted by (say) mild mixing, it is possible to explain the
difference in the abundance patterns by a different set of initial
angular momenta.  Such a difference could arise in the context of the
currently popular model for the origin of the range of rotation rates,
namely that the lifetime of accretion disks determines the rotation
rate, if globular cluster stars experienced more frequent interactions
which disrupted their accretion disks early in their lifetimes
relative to the lower density systems that the halo stars arose from.
It is more challenging, however, to explain why stars with similar
thermal structures should experience completely different depletion
histories.

The {\it complete} absence of depletion in Population II stars would also
have to be reconciled with the strong evidence for depletion in
Population I stars which is not predicted by standard models.

At the same time, it is also clear that none of the existing
theoretical models provide a complete description of the complex
lithium abundance pattern seen in stars.  Although the most recent
classes of rotational models are 
reasonably successful at reproducing the observed angular momentum
evolution of low mass Population I stars, they do not reproduce the
solar rotation profile as inferred from helioseismology.  They also
require an extrapolation of the initial conditions from Population I
to Population II stars, which may introduce systematic errors in the 
calculations.  Further theoretical work is clearly needed, and a more
refined set of models could potentially alter the inferred degree of
stellar depletion.

\section{Summary}

We have compared theoretical models with the observational Population
II lithium data to obtain bounds on stellar lithium depletion.  From a
combination of the dispersion in the data, the detection of $Li^6$,
and the flatness of the halo plateau interesting bounds can be set on
the stellar depletion of lithium in Population II stars.  The majority
of tests are roughly consistent with depletion at the 0.15-0.2 dex
level, with a firm upper bound of 0.5-0.6 dex from a combination of the
detected $Li^6$ abundance in HD 84937 and an extreme chemical
evolution model.  The most recent data set of RNB places the most
severe observational constraints on the dispersion of lithium in halo
stars.  It is marginally consistent with the least depleted set of
models of PWSN including rotational mixing; a larger statistical
sample would permit a more definitive test of the limits on depletion
from mixing in halo stars.  There are unexplained differences between 
the equivalent width measurements of Thorburn (1994) and RNB which need to be
understood; indeed, the systematic differences between investigators
on the absolute level of the plateau are approaching the uncertainty
in the inferred degree of stellar depletion.  

A new generation of theoretical models is needed to further refine 
our understanding of light element depletion in stars.  
The interaction between different physical mechanisms, such as 
microscopic diffusion, mass loss, and rotational mixing, 
and a better physical description of each of them, 
may prove important in this context.

Finally, any stellar lithium depletion has implications for BBN.  PWSN
discussed the implications of a higher primordial lithium abundance;
see Olive (these proceedings) for the case of negligible stellar
depletion.  If the preliminary data from the BOOMERANG mission is
confirmed (Lange et al. 2000), we note that there may be a
disagreement between the stellar lithium abundances and the
predictions of standard BBN as well as the low deuterium results of
Tytler (these proceedings.)  This is significantly relaxed if stellar
lithium depletion has occured.

\section{Acknowledgements}

M.P. would like to acknowledge support from NASA grant NAG5-7150 and
NSF grant AST-9731621. C.C. thanks the Action Sp\'ecifique de Physique 
Stellaire and the Conseil National Fran\c cais d'Astronomie for support.
%CC1906
C.P.D. acknowledges support from the United States
National Science Foundation under grant AST-9812735.
%CC1906

\end{document}